\documentclass[paper,prd,twocolumn,nofootinbib,preprintnumbers,amsmath,amssymb]{revtex4-2}

\usepackage{latexsym,graphicx,amssymb,amsmath,mathrsfs,bbold} 
\usepackage{slashed,mathbbol}
\usepackage[utf8]{inputenc}
\usepackage{color}
\usepackage{makecell}

\newcommand{\be}{\begin{equation}}      
\newcommand{\ee}{\end{equation}}      
\newcommand{\bea}{\begin{eqnarray}}      
\newcommand{\eea}{\end{eqnarray}}    
\newcommand{\Tr}{\,\textrm{Tr}\,}

\makeatletter
\renewcommand\appendix{\par
\setcounter{section}{0}%   
\setcounter{subsection}{0}% 
\gdef\thesection{\appendixname\space\@Alph\c@section}}

\long\def\unmarkedfootnote#1{{\long\def\@makefntext##1{##1}\footnotetext{#1}}}
\makeatother

\begin{document} 

\title{Chiral symmetry restoration in QC$_2$D from effective model \\using the functional renormalization group}
\author{G. Fej\H{o}s}
\affiliation{Institute of Physics and Astronomy, E\"otv\"os University, 1117 Budapest, Hungary,}
\affiliation{Interdisciplinary Theoretical and Mathematical Sciences Program (iTHEMS), RIKEN, Wako, Saitama 351-0198, Japan}
\author{D. Suenaga}
\affiliation{Kobayashi-Maskawa Institute for the Origin of Particles and the Universe, Nagoya University, Nagoya 464-8602, Japan}
\affiliation{Research Center for Nuclear Physics, Osaka University, Ibaraki 567-0048, Japan}

\begin{abstract}
The order of the chiral phase transition in two-color and two-flavor QC$_2$D is investigated using the functional renormalization group (FRG) technique in an effective model setting. We calculate the $\beta$ function of all couplings in the dimensionally reduced Ginzburg-Landau free energy functional with Pauli-Gürsey SU(4) symmetry. We compare results of the perturbative $\epsilon$ expansion approach with those obtained via the FRG, evaluated directly in $d=3$ dimensions. The perturbative results suggest that the fixed-point structure is more intricate than that of three-color QCD, a conclusion further supported by the FRG analysis. Both methods display an infrared stable $O(6)$ fixed point at infinite axial anomaly; however, the FRG approach also reveals the existence of $U_A(1)$ anomaly-free fixed points, which can become infrared stable if the anomaly in the underlying theory vanishes at the critical temperature. These findings imply that the phase transition can be of second order, consistent with earlier findings for three colors.
\end{abstract}

\maketitle

\section{Introduction}

In two-color quantum chromodynamics (QC$_2$D), it is well known that the so-called sign problem in Monte Carlo simulations disappears even in the presence of a nonzero quark-number chemical potential, in contrast to three-color QCD~\cite{nagata22}. Hence, first-principles lattice computations are applicable at finite density: therefore, QC$_2$D is regarded as a useful testing ground for elucidating a wide regime of the phase diagram, particularly the cold and dense regime, and has attracted significant attention in recent times \cite{boz13,Braguta2016,buividovich20,astrakhantsev20,braguta23,iida24,kogut00,ratti04,sun07,khan15,suenaga22}. While QC$_2$D does not display the sign problem, it does share a key property with QCD: at low temperatures and/or densities, chiral symmetry is spontaneously broken. To achieve a satisfactory understanding of the strongly interacting matter in QC$_2$D, it is important to identify the similarities and differences in chiral symmetry breaking compared to ordinary QCD. In this paper, our aim is to provide new insights into the order of the phase transition leading to the spontaneous breaking of chiral symmetry for two colors ($N_c=2$) and two flavors ($N_f=2$) in the zero quark mass limit.

First, one notes that for $N_c=2$, diquarks form a color singlet and are regarded as baryons. As such, they belong to the same multiplet as the hadronic composites. Consequently, instead of the usual two-flavor $SU(2)\times SU(2)$ chiral symmetry, the system exhibits an emergent $SU(4)$ Pauli-Gürsey symmetry~\cite{kogut00}. The construction of the aforementioned multiplet, including all parity partners, as well as the formulation of a corresponding linear sigma model (L$\sigma$M), can be found in \cite{suenaga22,suenaga25}. Here, we note only that if one is interested solely in the possibility of a continuous phase transition and the system's critical behavior, it suffices to consider the dimensionally reduced L$\sigma$M [also known as the Ginzburg–Landau (GL) theory for the free energy functional] and search for renormalization group fixed points that could correspond to a thermal phase transition.

This has already been worked out for $N_c=3$, but the order of the transition in the chiral limit remains a subject of debate. According to the traditional scenario based on the $\epsilon$ expansion of renormalization group flows, the transition is first order for any number of flavors, with a possible exception in the $N_f=2$ case, where a strong axial anomaly might induce a second-order transition \cite{pisarski84}. However, recent evidence increasingly suggests that the transition may be second order regardless of the number of flavors, presumably up to the conformal window \cite{cuteri21,dini21,zhang24,bernhardt23,kousvos22,giacosa25}. Although the corresponding renormalization group fixed point cannot be accessed using the $\epsilon$ expansion, the functional renormalization group (FRG) framework provides clear indications of its existence \cite{fejos22,fejos24}.

No such analyses, including the $\epsilon$ expansion, have been conducted for QC$_2$D so far. Similar to the original $N_c=3$ argument, the absence of an infrared-stable fixed point at the critical temperature would suggest that the system cannot exhibit scaling behavior. Therefore, if the transition does indeed exist, it cannot be second order and is presumably first order. Following a similar line of reasoning, in this paper, we aim to investigate the renormalization group flows of the GL free energy functional for QC$_2$D with $N_f=2$ to search for infrared-stable fixed points that could correspond to a second-order transition. Our approach is twofold: on the one hand, we apply the traditional $\epsilon$ expansion around $d=4$, ultimately extrapolating the $\beta$ functions to $d=3$. On the other hand, we utilize the FRG technique to make predictions directly in $d=3$ without relying on perturbation theory.

The paper is organized as follows. In Sec. II, we define the model and present our ansatz for the free energy. In Sec. III, we outline the procedure for obtaining the scale dependencies (i.e., the $\beta$ functions) of each coupling using the FRG formalism. We first examine the fixed-point structure through the $\epsilon$ expansion and then evaluate the flows directly in $d=3$ using the FRG. Finally, Sec. IV summarizes our findings.

\section{Effective model}

The order parameter in the GL free energy for $N_c=2$, $N_f=2$ QCD is a $4\times 4$ matrix, belonging to a subset of the $U(4)$ algebra in the following sense~\cite{suenaga22,suenaga25}:
\bea
\label{Eq:Sigma}
\Sigma = (S_a - i P_a) \hat{X}^a \hat{E} + (B'_i - iB_i)\hat{X}^i\hat{E},
\eea
where $\hat{E}$ is the $4\times 4$ symplectic matrix,
\bea
\hat{E} = \begin{pmatrix} 0 & {\bf 1}_{2\times 2} \\ -{\bf 1}_{2\times 2} & 0 \end{pmatrix},
\eea
while
\bea
\hat{X}^{a=0,1,2,3}&=& \frac{1}{2\sqrt2} \begin{pmatrix} \tau^a & 0 \\ 0 & (\tau^a)^T \end{pmatrix}, \nonumber\\
\hat{X}^{i=4,5} &=& \frac{1}{2\sqrt2} \begin{pmatrix} 0 & D^i \\ (D^i)^\dagger & 0 \end{pmatrix}.
\eea
Here $\tau^0 = {\bf 1}$, $\tau^{a=1,2,3}$ are the Pauli matrices, and $D^4 = \tau^2$, $D^5 = i\tau^2$. We stress that, in (\ref{Eq:Sigma}) all parity partners are included; physically $S_a$ and $P_a$ represent scalar and pseudoscalar mesons while $B_i'$ and $B$ represent negative-parity and positive-parity diquark baryons, resulting in a total of 12 degrees of freedom. Note that without all parity partners introduced appropriately, we found that the RG equations did not close with respect to couplings of the UV free energy.

Introducing the dual tensor of $\Sigma$, 
\bea
\tilde{\Sigma}_{ij} =\frac12\epsilon_{ijkl} \Sigma_{kl}, 
\eea
the ultraviolet (i.e., fluctuationless) free energy is as follows:
\begin{widetext}
\bea
\label{Eq:Lag}
{\cal F}_{\Lambda}&=&\int d^3x \Big[ \Tr [\partial_{\mu} \Sigma^\dagger \partial^{\mu} \Sigma] + m^2\Tr [\Sigma^\dagger \Sigma] + \tilde{g}_1 \big(\Tr [\Sigma^\dagger \Sigma]\big)^2 + g_2 \Tr [\Sigma^\dagger \Sigma\Sigma^\dagger \Sigma]\nonumber\\
&&\hspace{1cm}+\tilde{b}_1 \big(\Tr [\Sigma^\dagger \Sigma] \big)^3+c_1 \Tr [\Sigma^\dagger \Sigma] \Tr [\Sigma^\dagger \Sigma\Sigma^\dagger \Sigma] \nonumber\\
&&\hspace{1cm}+\frac{a_2}{2}\Tr [\tilde{\Sigma}\Sigma + \tilde{\Sigma}^\dagger \Sigma^\dagger] + \frac{a_4}{4}\big(\Tr[\tilde{\Sigma}\Sigma + \tilde{\Sigma}^\dagger \Sigma^\dagger] \big)^2 + \frac{a_{22}}{2}\Tr [\Sigma^\dagger \Sigma]\Tr[\tilde{\Sigma}\Sigma + \tilde{\Sigma}^\dagger \Sigma^\dagger]\nonumber\\
&&\hspace{1cm}+\frac{\tilde{b}_2}{2} (\Tr [\Sigma^\dagger \Sigma])^2 \Tr[\tilde{\Sigma}\Sigma + \tilde{\Sigma}^\dagger \Sigma^\dagger]+\frac{b_3}{4} \Tr [\Sigma^\dagger \Sigma] \big(\Tr[\tilde{\Sigma}\Sigma + \tilde{\Sigma}^\dagger \Sigma^\dagger]\big)^2+\frac{b_4}{8} \big(\Tr[\tilde{\Sigma}\Sigma + \tilde{\Sigma}^\dagger \Sigma^\dagger]\big)^3\nonumber\\
&&\hspace{1cm}+\frac{c_2}{2} \Tr [\Sigma^\dagger \Sigma\Sigma^\dagger \Sigma] \Tr[\tilde{\Sigma}\Sigma + \tilde{\Sigma}^\dagger \Sigma^\dagger]\Big],
\eea
\end{widetext}
where the first two lines in the right-hand side are fully symmetric under chiral rotations, while the remaining terms break $U_A(1)$ transformations\footnote{Throughout this paper, we occasionally use chiral symmetry to refer to Pauli-G\"{u}rsey symmetry.}. Note that in contrast to the $N_c=3$ system, the standard Kobayashi-Maskawa-'t Hooft determinant is not an independent invariant:
\bea
\det \Sigma + \det \Sigma^\dagger &=& \frac{1}{16} \big(\Tr [\tilde{\Sigma}\Sigma + \tilde{\Sigma}^\dagger \Sigma^\dagger]\big)^2 +\frac12 \Tr [\Sigma^\dagger \Sigma \Sigma^\dagger \Sigma]\nonumber\\
&-&\frac14 \big(\Tr [\Sigma^\dagger \Sigma])^2,
\eea
therefore, it is excluded from (\ref{Eq:Lag}). This also means that the only way to incorporate the $U_A(1)$ anomaly into the system is through the operator $\sim \Tr [\tilde{\Sigma}\Sigma + \tilde{\Sigma}^\dagger \Sigma^\dagger]$. That is to say, the form of (\ref{Eq:Lag}) is the most general expression allowed by symmetry and (perturbative) renormalizability in $d=3$\footnote{In $d=3$, dimensionless couplings correspond to ${\cal O}(\Sigma^6)$ terms.}. We also note that this is the first FRG study that takes into account the $U_A(1)$ anomaly appropriately for $N_c=2$. 

The full free energy functional, which incorporates all fluctuations, can only depend on combinations of $\Sigma$ that are invariant under chiral rotations and potentially break $U_A(1)$. The following set represents a convenient (but nonunique) choice of independent invariants:
\begin{subequations}
\label{Eq:inv}
\bea
I_1 &=& \Tr [\Sigma^\dagger \Sigma],  \\
I_2 &=& \Tr [(\Sigma^\dagger \Sigma - \frac14\Tr[\Sigma^\dagger \Sigma]\mathbb{1} )^2], \\
I_4 &=&  \Tr [(\Sigma^\dagger \Sigma - \frac14\Tr[\Sigma^\dagger \Sigma]\mathbb{1} )^4], \\
I_A &=& \frac12\Tr [\tilde{\Sigma}\Sigma + \tilde{\Sigma}^\dagger \Sigma^\dagger].
\eea
\end{subequations}
Any other chiral invariant [including those that break $U_A(1)$] can be expressed in terms of Eqs. (\ref{Eq:inv}). Note that the combination $\Tr [(\Sigma^\dagger \Sigma - \frac14\Tr[\Sigma^\dagger \Sigma]\mathbb{1} )^3]$ is identically zero\footnote{In principle, one has $\Tr [(\Sigma^\dagger \Sigma - \frac14\Tr[\Sigma^\dagger \Sigma]\mathbb{1} )^{2n+1}]\equiv 0$.}, and thus it cannot appear in the free energy.

In order to decide whether the system can show critical behavior and thus undergo a second-order transition, one needs to know whether the free energy can exhibit scaling behavior. In this study, the scale-dependent free energy, ${\cal F}_k$, is approximated as
\bea
\label{Eq:ansatz}
{\cal F}_k = \int_x \Big( \Tr [\partial_i \Sigma^{\dagger} \partial_i \Sigma] + V_k(\Sigma)\Big),
\eea
where $V_k$ is the scale-dependent effective potential, in which only (perturbatively) relevant and marginal interactions are kept:
\bea
\label{Eq:Vk}
V_k &=& m_k^2 I_1 + g_{1,k} I_1^2 + g_{2,k} I_2 + b_{1,k}I_1^3+c_{1,k} I_1 I_2\nonumber\\
&+&a_{2,k} I_A + a_{4,k} I_A^2 + a_{22,k} I_1 I_A \nonumber\\
&+&b_{2,k} I_1^2I_A + b_{3,k}I_1I_A^2+b_{4,k}I_A^3+c_{2,k}I_2 I_A.
\eea
The form of (\ref{Eq:ansatz}) is sometimes called the local potential approximation (LPA). Note that, the structure of (\ref{Eq:Vk}) matches that of (\ref{Eq:Lag}) with the substitutions $g_1=\tilde{g}_1 + g_2/4$, $b_1=\tilde{b}_1 + c_1/4$, and $b_2=\tilde{b}_2 + c_2/4$. Also, note that, $k$ indicates the scale of separation between IR and UV modes, which, in the FRG formulation, is done via an appropriately chosen regulator term. The reader is referred to \cite{wetterich93,berges02,kopietz10} for details of the formalism.

\section{Renormalization group equations and fixed-point analyses}

The scale-dependent free energy satisfies the Wetterich equation \cite{wetterich93}:
\bea
\label{Eq:flowFk}
k\partial_k {\cal F}_k = \frac12 \int \Tr[k\partial_k R_k ({\cal F}_k'' + R_k)^{-1}],
\eea
where ${\cal F}_k''$ is the second functional derivative matrix of ${\cal F}_k$ with respect to all dynamical variables $S^a, P^a, B'^i, B^i$. In this study, we employ the optimal regulator for the LPA \cite{litim01}, i.e., $R_k(q)=(k^2-{\bf q}^2)\Theta(k^2-{\bf q}^2)$. This leads to the following flow equation for the effective potential:
\bea
\label{Eq:flowVk}
k\partial_k V_k = \frac{k^{d+2}}{d}\Omega_d \Tr (k^2 + V_k'')^{-1},
\eea
where $\Omega_d\equiv \int d\Omega/(2\pi)^d$ comes from the angular integrals in the $d$ dimensional Euclidean space, while $V_k''$ is the second derivative matrix of $V_k$:
\bea
\label{Eq:massmatrix}
V_k'' = \begin{pmatrix}
\frac{\partial^2 V_k}{\partial S_i \partial S_j} & \frac{\partial^2 V_k}{\partial S_i \partial B_j} & \frac{\partial^2 V_k}{\partial S_i \partial P_j} & \frac{\partial^2 V_k}{\partial S_i \partial B'_j} \\
\frac{\partial^2 V_k}{\partial B_i \partial S_j} & \frac{\partial^2 V_k}{\partial B_i \partial B_j} & \frac{\partial^2 V_k}{\partial B_i \partial P_j} & \frac{\partial^2 V_k}{\partial B_i \partial B'_j} \\
\frac{\partial^2 V_k}{\partial P_i \partial S_j} & \frac{\partial^2 V_k}{\partial P_i \partial B_j} & \frac{\partial^2 V_k}{\partial P_i \partial P_j} & \frac{\partial^2 V_k}{\partial P_i \partial B'_j} \\
\frac{\partial^2 V_k}{\partial B'_i \partial S_j} & \frac{\partial^2 V_k}{\partial B'_i \partial B_j} & \frac{\partial^2 V_k}{\partial B'_i \partial P_j} & \frac{\partial^2 V_k}{\partial B'_i \partial B'_j} \\
\end{pmatrix}.
\eea
Note that $V_k$ is considered to be a function of chiral invariants; therefore, when evaluating (\ref{Eq:massmatrix}) one differentiates with respect to the invariants first, and then takes the derivatives of the invariants with respect to the fields. After plugging the ansatz (\ref{Eq:Vk}) into (\ref{Eq:flowVk}), we get, by definition, on the left-hand side the sum of the logarithmic scale derivatives of each coupling, multiplied by their respective operators. The right-hand side must be compatible with this structure, i.e., after taking the inverse and expanding the expression with respect to the components of the background field, the same operator structure must emerge. This allows us to identify, at each order, the corresponding flows of the couplings. Note that one does not need to calculate the inverse matrix of $k^2 + V_k''$ in the most general ($2 \times 6 = 12$ component) background field; it is sufficient to consider, at each order, such a background, which allows for the unique separation of each invariant that appears at that given order. We tried several different background fields and always obtained the same results. It is important to emphasize that the structure dictated by (\ref{Eq:Vk}) is, therefore, indeed inherited by the right-hand side of the flow equation (\ref{Eq:flowVk}), which means that all independent invariants have been properly included in the effective potential.

Then, the $\beta$ function of some $\lambda^{(n)}$ coupling that corresponds to an operator containing $n$ scalar fields is defined as follows:
\bea
\beta_{\lambda^{(n)}} &=& k\partial_k \bar{\lambda}^{(n)}_k\nonumber\\
&\equiv&-\big(n+\frac{d}{2}(2-n)\big) \bar{\lambda}^{(n)}_k + k\partial_k \lambda^{(n)}_k k^{-n-\frac{d}{2}(2-n)}, \nonumber\\
\eea
where $\bar{\lambda}_k^{(n)}\equiv \lambda_k^{(n)}k^{-n-\frac{d}{2}(2-n)}$ is the dimensionless coupling. First, we present the results for the $\beta$ functions in $d = 4 - \epsilon$ dimensions at the leading order in $\epsilon$, and then discuss them as evaluated directly in $d = 3$ via the FRG.

\subsection{$\epsilon$ expansion}

To the best of our knowledge, the $\beta$ functions of the model defined by (\ref{Eq:Lag}) in the $\epsilon$ expansion have not yet been calculated. Even though our main goal in this study is to perform a direct $d = 3$ computation within the framework of the FRG, for the sake of completeness, we also wish to include the results of the $\epsilon$ expansion.

One can obtain these standard perturbative expressions from either the textbook Wilsonian picture or from the field theoretical RG, but it has also been known for some time that they can be readily obtained using the FRG, i.e., from Eq. (\ref{Eq:flowVk}) \cite{fejos14}. We proceed with the latter.

Close to $d = 4$, the ${\cal O}(\Sigma^6)$ terms are irrelevant; therefore, we may set $b_1 \equiv b_2 \equiv b_3 \equiv b_4 \equiv c_1 \equiv c_2 \equiv 0$, which leaves us with six flowing parameters. Identifying interacting fixed points is similar to that in the $O(N)$ model. We do not list the exact form of the $\beta$ functions that arise from (\ref{Eq:flowVk}), on the one hand due to their complexity, and on the other hand because, close to $d = 4$, we do not need them. The reason is the following. One notices that the flows of the quartic couplings ($g_1, g_2, a_4, a_{22}$) have the schematic structure of
\bea
\beta \sim -\epsilon \times \mathrm{coupling} + \hspace{0.05cm}\#\hspace{0.1cm}(\mathrm{coupling})^2,
\eea
where $\epsilon = 4 - d$, meaning that in order for the perturbation theory to make sense, each coupling must be of ${\cal O}(\epsilon)$ at any nontrivial fixed point. This also implies that, at such a fixed point, the quadratic couplings ($m^2, a_2$) are of ${\cal O}(\epsilon)$ as well, since for them one has schematically
\begin{subequations}
\bea
\beta_{m^2} &\sim& -2 \bar{m}_k^2 +\hspace{0.05cm}\#\hspace{0.1cm}\mathrm{coupling}, \\
\beta_{a_2} &\sim& -2 \bar{a}_{2,k} + \hspace{0.05cm}\#\hspace{0.1cm}\mathrm{coupling}.
\eea
\end{subequations}
That is, when calculating the position of a fixed point, both $\bar{m}^2$ and $\bar{a}_2$ can be dropped in the propagators appearing in the $\#$ factors of each expression of a $\beta$ function, since they produce subleading contributions in $\epsilon$. Close to a nontrivial fixed point, therefore, at leading order, we arrive at the following $\beta$ functions:
\begin{subequations}
\label{Eq:betaeps}
\bea
\label{Eq:betaepsm2}
\beta_{m^2} &=& -2 \bar{m}_k^2 - 7\bar{g}_{1,k}-\frac{5}{4}\bar{g}_{2,k} - \bar{a}_{4,k}, \\
\label{Eq:betaepsa2}
\beta_{a_2} &=& -2 \bar{a}_{2,k} - 4\bar{a}_{22,k}, \\
\label{Eq:betaepsg1}
\beta_{g_1} &=& -\epsilon \bar{g}_{1,k} + 20\bar{g}^2_{1,k}+5\bar{g}_{1,k}\bar{g}_{2,k}+\frac{5}{4}\bar{g}_{2,k}^2\nonumber\\
&&+4\bar{a}_{4,k}^2+7\bar{a}_{22,k}+4\bar{g}_{1,k}\bar{a}_{4,k}, \\
\label{Eq:betaepsg2}
\beta_{g_2} &=& -\epsilon \bar{g}_{2,k} +12\bar{g}_{1,k}\bar{g}_{2,k}-4\bar{a}_{4,k}\bar{g}_{2,k},\\
\label{Eq:betaepsa4}
\beta_{a_4} &=& -\epsilon \bar{a}_{4,k} +16\bar{a}_{4,k}^2+7\bar{a}^2_{22,k}\nonumber\\
&&+12\bar{a}_{4,k}\bar{g}_{1,k}-5\bar{g}_{2,k}\bar{a}_{4,k},\\
\label{Eq:betaepsa11}
\beta_{a_{22}} &=& -\epsilon \bar{a}_{22,k} +28\bar{a}_{22,k}(\bar{g}_{1,k}+\bar{a}_{4,k}),
\eea
\end{subequations}
where we formally set $\Omega_d \rightarrow 1$.\footnote{This can always be achieved by rescaling the quartic couplings with $\Omega_d$.} We note that Eqs. (\ref{Eq:betaepsg1}), (\ref{Eq:betaepsg2}), (\ref{Eq:betaepsa4}), and (\ref{Eq:betaepsa11}) do not depend on the choice of the regulator. This is because, close to $d=4$, the flows of the quartic couplings are regulator independent at the leading order, provided that the quadratic coupling(s) are set to zero \cite{berges02}. Equations (\ref{Eq:betaepsm2}) and (\ref{Eq:betaepsa2}) are regulator dependent, but it turns out that since the fixed-point equations for the quartic couplings decouple from those of the quadratic ones, the latter have no influence on the scaling dimensions around the fixed point(s). Therefore, the physical predictions of the $\epsilon$ expansion at leading order do not depend on the regularization scheme. This is why the same result can also be obtained via the field theoretical RG.

The fixed points corresponding to the zeros of Eqs. (\ref{Eq:betaeps}) are summarized in Table I for $\epsilon = 1$ ($d \to 3$). Since Eqs. (\ref{Eq:betaeps}) are invariant under the simultaneous sign flips $\bar{a}_2 \to -\bar{a}_2$ and $\bar{a}_{22} \to -\bar{a}_{22}$, we can choose $\bar{a}_2 < 0$. Apart from the trivial Gaussian fixed point, we find an $O(12)$ fixed point where all the anomalous couplings $\bar{a}_2$, $\bar{a}_4$, $\bar{a}_{22}$, and the quartic coupling $\bar{g}_2$ vanish. [Note that $\bar{g}_2$ is the coupling that breaks the $O(12)$ invariance of an anomaly-free theory.] Additionally, we find three fixed points (A1, A2, A3) with nonvanishing anomaly, where $\bar{g}_2 = 0$, and two more (A4, A5) with $\bar{g}_2 \neq 0$, and the anomaly is also present.

By diagonalizing the stability matrix $\partial_j \beta_i$,\footnote{Here we differentiate all $\beta$ functions with respect to all couplings.} we find that none of the fixed points has exactly one relevant direction. For the Gaussian and $O(12)$ fixed points, one might be tempted to investigate the system without anomalous couplings, as these fixed points are free from anomaly couplings and thus exist in a theory without the $U_A(1)$ anomaly. In such a case, the number of coupling directions decreases by three. However, even then, none of the fixed points has exactly one relevant direction. This leads to the conclusion that, similar to the $N_c = 3$, $N_f = 2$ case, the $\epsilon$ expansion suggests that the chiral phase transition cannot be of second order.

There is one important caveat before drawing conclusions from the $\epsilon$ expansion. In comparison with the $N_c = 3$, $N_f = 2$ case, it is crucial to discuss a scenario where the anomaly is very strong in the sense that $m^2 - a_2 \to \infty$, but the combination $m^2 + a_2$ remains finite. In such a case, half of the modes become infinitely heavy, meaning that they can effectively be integrated out. This, in effect, yields all quartic invariants to become the square of the squared sum of the remaining fields, implying that the theory is indeed $O(6)$ symmetric. For the $N_c = 3$, $N_f=2$ model, identical assumptions lead to the existence of an infrared-stable $O(4)$ fixed point at the critical temperature \cite{grahl13}. Similarly, the $O(6)$ model also possesses an infrared-stable fixed point at the critical temperature. Therefore, a very large anomaly may still cause a second-order transition for $N_c=2$, $N_f=2$, but with $O(6)$ critical exponents. This fixed point is indicated in the last line of Table I.

\begin{table}[t]
\vspace{0.2cm}
  \begin{tabular}{ c | c | c | c | c | c | c | c | c }
   FP & $\bar{m}^2$ & $\bar{a}_2$ & $\bar{g}_1$ & $\bar{g}_2$ & $\bar{a}_4$ & $\bar{a}_{22}$ & {\scriptsize RD} & {\scriptsize RD} {\tiny w/o $U_A(1)$} \\ \hline\hline
   Gauss & 0 & 0 & 0 & 0 & 0 & 0 & {\bf 6} & {\bf 3} \\ \hline
   O(12) & -7/40 & 0 & 1/20 & 0 & 0 & 0 & {\bf 4} & {\bf 2} \\ \hline
   A1 & -1/7 & 0 & 1/28 & 0 & 1/28 & 0 & {\bf 3} & {\scriptsize N/A} \\ \hline
   A2 & -9/88 & 0 & 1/44 & 0 & 1/22 & 0 & {\bf 3} & {\scriptsize N/A} \\ \hline
   A3 & -1/14 & -1/14 & 1/56 & 0 & 1/56 & 1/28 & {\bf 5} & {\scriptsize N/A}  \\ \hline
   A4 & -9/88 & 0 & 3/44 & -2/11 & -1/22 & 0 & {\bf 5} & {\scriptsize N/A} \\ \hline
   A5 & -1/7 & 0 & 1/14 & -1/7 & -1/28 & 0 & {\bf 4} & {\scriptsize N/A} \\ \hline
   O(6) & -1/7 & $\infty$ & 1/14 & {\scriptsize N/A} & {\scriptsize N/A} & {\scriptsize N/A} & {\bf 1} & {\scriptsize N/A}  \\ \hline
  \end{tabular}
  \caption{Fixed points in the $\epsilon$ expansion extrapolated to $d=3$, with their respective number of relevant directions (RD). The number of the RDs with and without the $U_A(1)$ anomaly are indicated separately. Note that, in case of the $O(6)$ fixed point, the mass parameter is shifted as $m^2+a_2\rightarrow m^2$.}
\end{table}

\begin{table*}[t]
  \begin{tabular}{ c | c | c | c | c | c | c | c | c }
   FP & $\bar{m}^2$ & $\bar{a}_2$ & $\bar{g}_1$ & $\bar{g}_2$ & $\bar{a}_4$ & $\bar{a}_{22}$ & {\scriptsize RD} & {\scriptsize RD} {\tiny w/o $U_A(1)$}\\ \hline\hline
   Gauss & 0 & 0 & 0 & 0 & 0 & 0 & {\bf 6} & {\bf 3} \\ \hline
   O(12) & -0.2964 & 0 & 0.0314 & 0 & 0 & 0 & {\bf 4} & {\bf 2} \\ \hline
   A1 & -0.25 & 0 & 0.0264 & 0 & 0.0264 & 0 & {\bf 3} & {\scriptsize N/A} \\ \hline
   A2 & -0.3334 & 0 & 0.0351 & 0 & -0.0232 & 0 & {\bf 4} & {\scriptsize N/A} \\ \hline
   A2b & -0.3334 & 0 & 0.0119 & 0.0927 & 0.0232 & 0 & {\bf 3} & {\scriptsize N/A} \\ \hline
   A3 & -1/8 & -1/8 & 27/2048 & 0 & 27/2048 & 27/1024 & {\bf 4} & {\scriptsize N/A} \\ \hline
   A4 & -0.25 & 0 & 0.0527 & -0.1055 & -0.0264 & 0 & {\bf 3} & {\scriptsize N/A} \\ \hline
   A5 & -0.3162 & 0 & 0.0457 & -0.0516 & -0.0335 & 0 & {\bf 3} & {\scriptsize N/A} \\ \hline
   A5b & -0.3162 & 0 & 0.0122 & 0.0823 & 0.0335 & 0 & {\bf 2} & {\scriptsize N/A} \\ \hline
   AF$_1$ & -0.3433 & 0 & 0.0126 & 0.1070 & 0 & 0 & {\bf 3} & {\bf 1} \\ \hline
   AF$_2$ & -0.2249 & 0 & 0.0541 & -0.1406 & 0 & 0 & {\bf 3} & {\bf 1} \\ \hline
   O(6) & -1/4 & $\infty$ & 27/512 & {\scriptsize N/A} & {\scriptsize N/A} & {\scriptsize N/A} & {\bf 1} & {\scriptsize N/A} \\ \hline
  \end{tabular}
  \caption{Fixed points in the FRG method, evaluated directly in $d=3$, with their respective number of relevant directions (RD). The number of the RDs with and without the $U_A(1)$ anomaly are indicated separately. Note that, in case of the $O(6)$ fixed point, the mass parameter is shifted as $m^2+a_2\rightarrow m^2$.}
\end{table*}

\subsection{FRG treatment}

In this section, we discuss the renormalization group flows and fixed points directly in $d=3$. First, we note that although Eq. (\ref{Eq:flowVk}) may have been used to derive the results of the $\epsilon$ expansion, in principle, the FRG treatment is not restricted to working close to $d=4$. Equation (\ref{Eq:flowVk}) is nonperturbative in the sense that there is no need to assume the existence of any small parameter. It follows from Eq. (\ref{Eq:flowFk}) with the ansatz given by Eq. (\ref{Eq:ansatz}), which serves as the leading-order approximation of the derivative expansion. The latter has been shown to converge when using the optimal regulator defined above Eq. (\ref{Eq:flowVk}) \cite{balog19}, so evaluating the flows directly at $d=3$ without expanding in any small parameter is justified.

The calculation of each $\beta$ function follows the procedure outlined at the beginning of this section. The main differences between the direct $d=3$ calculation and its $d=4-\epsilon$ counterpart are as follows. In the $d=4-\epsilon$ case, the flows are relatively simple because we are working at leading order in $\epsilon$, whereas in $d=3$ one cannot make such simplifications, and the expressions for the $\beta$ functions are much more involved. Additionally, since in $d=3$ the marginal interactions are of ${\cal O}(\Sigma^6)$, the number of couplings in the model is 12 [as shown in Eq. (\ref{Eq:ansatz})]. We do not attempt to present the expression for each flow here, but the corresponding formulas can be found in the Supplemental Material.

The procedure for obtaining the fixed points is as follows. After formally setting $\Omega_d \rightarrow 1$\footnote{Here, one needs to rescale the quartic couplings with $\Omega_d$, and the sextic ones with $\Omega_d^2$.}, we observe that the fixed-point equations for the marginal couplings, namely $\bar{b}_1, \bar{b}_2, \bar{b}_3, \bar{b}_4, \bar{c}_1, \bar{c}_2$, can be solved analytically. The values of these marginal couplings depend on the relevant couplings, i.e., $\bar{m}^2, \bar{a}_2, \bar{g}_1, \bar{g}_2, \bar{a}_4, \bar{a}_{22}$. Once the analytical expressions for the marginal couplings are obtained, they are substituted into the $\beta$ functions of the relevant couplings. This results in six fixed-point equations, similar to the case of the $\epsilon$ expansion [see Eqs. (\ref{Eq:betaeps})], which are solved numerically, along with a stability analysis.

The resulting fixed-point structure is shown in Table II. This structure is quite similar to that of Table I, with two main differences. First, the A2 and A5 fixed points split into two, resulting in the creation of A2b and A5b. Second, and more importantly, we observe the emergence of two completely new, anomaly-free fixed points (AF$_1$ and AF$_2$), which were not accessible in the $\epsilon$ expansion. This resembles the $N_c=3$ case, where such a fixed point is infrared stable at the critical point if the $U_A(1)$ anomaly is absent \cite{fejos22,fejos24}. This is also the case in our current $N_c=2$ scenario. If the $U_A(1)$ anomaly is present at the critical point, for both AF$_1$ and AF$_2$ the number of relevant directions is three, leaving no possibility for an attractive fixed point at the transition point. However, if the $U_A(1)$ anomaly disappears at the critical point, the anomalous directions are no longer present, reducing the number of relevant directions for the AF fixed points to one. In this case, both of the AF fixed points can correspond to a continuous phase transition at finite temperature. Note, however, that, the sign of the $g_2$ coupling (and accordingly, that of $c_1$) is different for them. One notes that AF$_1$ realizes a potential that is bounded from below, however, AF$_2$ does not (see next subsection). That is, for our purposes, it is expected that the AF$_1$ fixed point can be a candidate for a second-order chiral phase transition, even though by improving the approximation, it might turn out that AF$_2$ also corresponds to a stable potential; therefore, one should not entirely rule out its significance. The flow chart corresponding to all nonanomalous fixed points is sketched in Fig. 1.

\begin{figure}
\includegraphics[bb = 0 0 1065 1090,scale=0.22,angle=0]{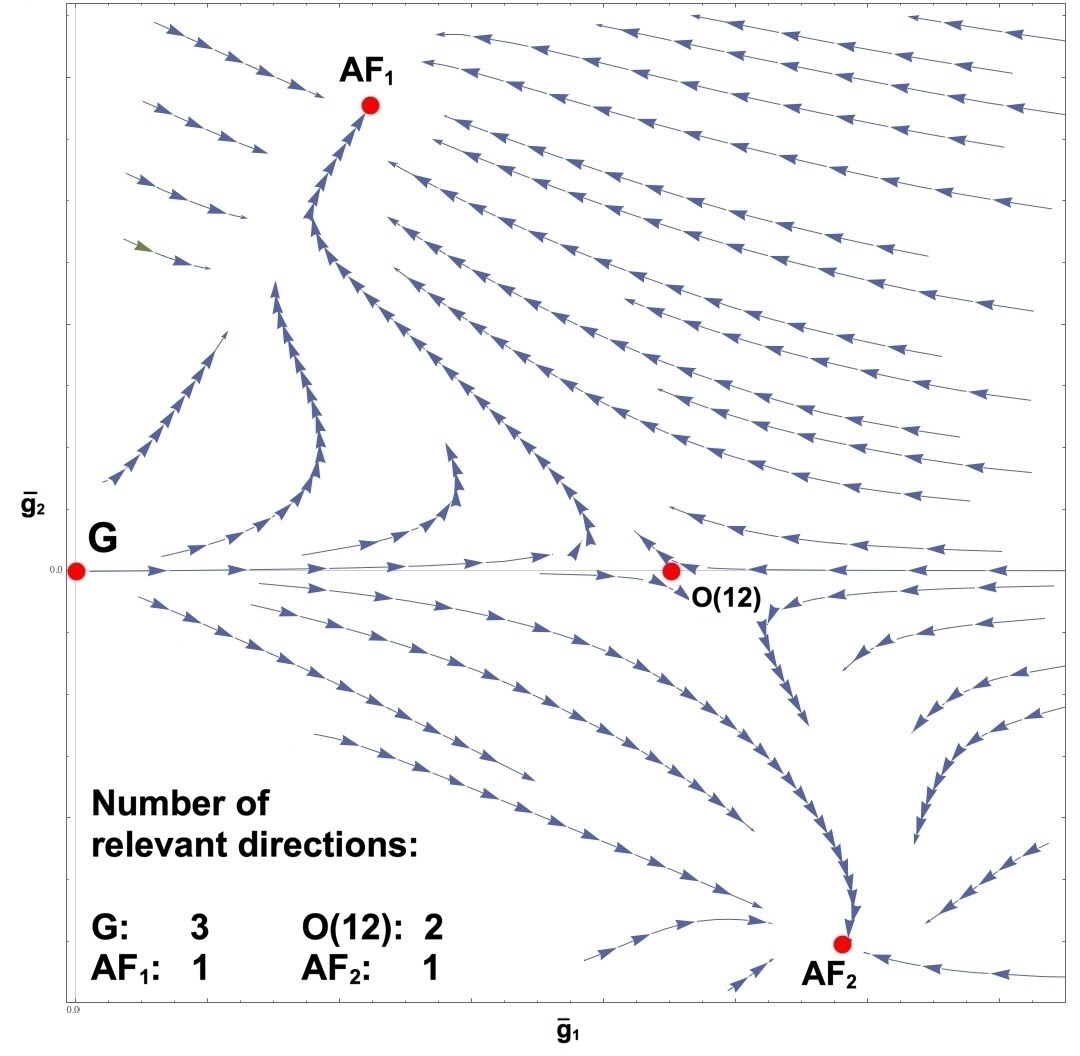}
\caption{Flow chart in case of the absence of the $U_A(1)$ anomaly. Note that, for the sake of visibility, the chart is shown for $\bar{m}^2\equiv 0$. This choice does not alter the essence of the flows, and makes it possible to show them in a 2D plot.}
\label{Fig:FP}
\end{figure}  

We also note that the analysis presented in the previous subsection regarding the $O(6)$ fixed point still holds within the FRG approach. The corresponding fixed point is shown in the last line of Table II. As a result, we identify two potential candidates for fixed points that could be responsible for a continuous phase transition, which aligns with the situation in the $N_c=3$, $N_f=2$ system. As discussed in the Introduction, for $N_c=3$, there is increasing evidence suggesting that the chiral transition is second order for zero quark masses. It would be interesting to determine whether the $N_c=2$ system exhibits the same property. Our current renormalization group analyses indicate that the transition could indeed be second order.

\subsection{Stability of the fixed-point potentials}

To determine whether the fixed points correspond to a potential that is bounded from below, we need to analyze the ${\cal O}(\Sigma^6)$ terms in (\ref{Eq:Vk}). Here, we only focus on the theory without the anomaly, in which case the stability condition for the potential is given by:
\bea
&&b_1^* I_1^3 + c_1^* I_1 I_2 \equiv\nonumber\\
&&\Big(b_1^* - \frac{c_1^*}{4}\Big) \Big(\Tr [\Sigma^\dagger \Sigma]\Big)^3 \!\!+c_1^* \Big(\Tr [\Sigma^\dagger \Sigma]\Big)^2\Tr [\Sigma^\dagger \Sigma\Sigma^\dagger \Sigma]>0,\nonumber\\
\eea
where $b_1^*$ and $c_1^*$ are the fixed-point values of the respective couplings. Since this condition remains invariant under the rescaling $\Sigma \rightarrow \lambda \Sigma$, we may set $\Tr [\Sigma^\dagger \Sigma]=1$. Furthermore, any unitary transformation of $\Sigma$ also preserves the condition, allowing us to consider $\Sigma$ as diagonal (and real). In this case, we obtain
\bea
\frac14 \leq \Tr[\Sigma^\dagger \Sigma\Sigma^\dagger \Sigma] \leq 1,
\eea
where the lower bound follows from the Cauchy--Schwarz inequality, while the upper bound is trivial and corresponds to a matrix with a single nonzero diagonal entry equal to one. Thus, the stability conditions reduce to:
\bea
\label{Eq:cond}
b_1^*>0, \quad b_1^*+3c_1^*/4>0,
\eea
which remain valid even after rescaling the couplings with $k$. Using the fixed-point equations for the $\bar{b}_1$ and $\bar{c}_1$ couplings, we express their values in terms of $\bar{m}^2$, $\bar{g}_1$, and $\bar{g}_2$. For the AF$_1$ fixed point, both conditions of (\ref{Eq:cond}) hold, whereas for AF$_2$, the second one is violated.

\section{Summary}

In this paper, we investigated whether the effective GL free energy functional for QC$_2$D with $N_f=2$ flavors can exhibit fixed points corresponding to a finite-temperature second-order phase transition. The free energy for $N_c=2$ differs from the $N_c=3$ case in that the theory displays an SU(4) Pauli-Gürsey symmetry instead of SU(2)$\times$SU(2). After constructing the most general renormalizable free energy, we calculated the $\beta$ functions of the couplings both in the $\epsilon$ expansion around $d=4$ and using the FRG approach, which allows for direct evaluation of scale dependencies in $d=3$.

Our findings using the $\epsilon$ expansion indicate that, despite the more intricate fixed-point structure found for $N_c=2$ compared to $N_c=3$, there is no infrared stable fixed point at the critical temperature for finite axial anomaly couplings, meaning no second-order transition can be associated. However, for an infinitely strong anomaly, an $O(6)$ fixed point, analogous to the $O(4)$ fixed point in the $N_c=3$ case, emerges and is infrared stable at the critical point.

The FRG method predicts a much richer structure. In particular, we identified the existence of two fixed points (labeled AF$_1$ and AF$_2$), where all anomalous couplings vanish. These fixed points are not infrared stable in a theory that breaks $U_A(1)$; however, in a GL functional that preserves $U_A(1)$, the stability analysis changes, and both fixed points are found to be infrared stable at the critical point. While AF$_2$ corresponds to an unphysical potential in our approximation, the AF$_1$ fixed point can be considered a candidate to describe a finite temperature second-order phase transition.

We showed that, similar to the $\epsilon$ expansion, the FRG approach also predicts an infrared stable $O(6)$ fixed point at infinite anomaly. Whether global renormalization group trajectories can reach the AF$_1$ or the $O(6)$ fixed point remains an open question, similar to the situation for $N_c=3$ \cite{fejos22,fejos24}. The answer may also depend on the QC$_2$D scale, which is a free parameter in the system. We also wish to emphasize that if the transition is indeed found to be of second order, determining the corresponding scaling exponents could provide indirect evidence of whether the anomaly is present at the critical point. If $O(6)$ exponents are observed, the anomaly must be present; however, if the exponents are consistent with those of the AF$_1$ fixed point, then the anomaly must vanish.

Exploring the chiral phase transition at finite temperature in QC$_2$D may provide valuable insights into $N_c$ dependencies. Although current QC$_2$D lattice simulations employ relatively heavy quark masses, it would be interesting to determine whether future simulations can detect any signs of a continuous phase transition in the chiral limit. Also, our analysis could be extended by investigating the role of anomalous dimensions at each fixed point and examining the influence of nonrenormalizable interactions in the GL functional, assessing their impact on the fixed-point structure. Additionally, the FRG method enables the determination of fully nonperturbative fixed-point potentials without assuming any specific functional form, potentially providing further evidence for the existence of infrared-stable fixed points. These aspects will be explored in future work and reported elsewhere.

\section*{Acknowledgments}

This work was supported by the Hungarian National Research, Development, and Innovation Fund under Project No. FK142594. G.F. would like to express his gratitude for the warm hospitality of the Kobayashi-Maskawa Institute for the Origin of Particles and the Universe at Nagoya University, where this research project was initiated. D.S. was supported by the JSPS KAKENHI Grants No. 23K03377 and No. 23H05439. D.S. expresses gratitude to the Young Researchers Overseas Program in Science of Nagoya University, which supported his stay at E\"{o}tv\"{o}s Lor\'{a}nd University.

\makeatletter
\@addtoreset{equation}{section}
\makeatother 
\renewcommand{\theequation}{A\arabic{equation}}

\end{document}